\documentclass[pra,twocolumn,superscriptaddress,10pt,noshowpacs]{revtex4}
\usepackage{pgf,pgfarrows,pgfnodes,pgfautomata,pgfheaps,pgfshade}
\usepackage{tikz}
\usepackage{tkz-euclide}
\usepackage{pgfplots}
\pgfplotsset{compat=1.18, width=9.5cm}
\usepackage{amsmath, amssymb}
\usepackage{bm}
\usepackage[utf8]{inputenc}
\usepackage{colortbl}
\usepackage{cancel}
\usepackage[T1]{fontenc}
\usepackage{xmpmulti}
\usepackage{animate}
\usepackage{verbatim}
\usepackage{graphicx,epstopdf}
\usepackage{bbm}
\usepackage{babel}
\usepackage{float}
\usepackage{gensymb}
\usepackage[pdftex,plainpages=false,colorlinks=true,citecolor=blue,linkcolor=blue,urlcolor=blue,filecolor=green,bookmarksopen=true]{hyperref}
\usepackage{amsfonts}
\usepackage{xcolor}
\usepackage{latexsym}
\usepackage{times,txfonts}
\usepackage{appendix}
\usepackage{caption}
\usepackage{subcaption}

\begin{document}
\title{Paradoxos da Relatividade}

\author{G. Alencar}
\affiliation{Departamento de Física, Universidade Federal do Ceará, 60440-554, Fortaleza, Ceará, Brazil}
\email{geova@fisica.ufc.br}

\author{J. Macedo}
\affiliation{Departamento de Física, Universidade Federal do Ceará, 60440-554, Fortaleza, Ceará, Brazil}
\email{geova@fisica.ufc.br}

\author{L. Maranhão}
\affiliation{Departamento de Física, Universidade Federal do Ceará, 60440-554, Fortaleza, Ceará, Brazil}
\email{leticiamaranhao@fisica.ufc.br}

\author{P. Carneiro}
\affiliation{Departamento de Física, Universidade Federal do Ceará, 60440-554, Fortaleza, Ceará, Brazil}
\email{patrick@fisica.ufc.br}

\date{\today}

\begin{abstract}
Nesse artigo, apresentamos vários paradoxos aparentes da relatividade restrita e suas respectivas soluções. Esse paradoxos aparecem desde o advento da relatividade, em 1905, e de fato nunca são paradoxos. Do ponto de vista didático, os paradoxos  são uma excelente ferramenta de aprendizado. Eles levam o estudante a confrontar, e abandonar,  vários conceitos centrais da teoria Galileana, como a simultaneidade e a rigidez dos corpos extensos. Particularmente, revisaremos uma nova e simples solução para o paradoxo dos gêmeos, encontrada recentemente por um dos presentes autores \cite{TwinAlencar}. Ela não necessita considerar referenciais acelerados ou sinais de luz, que são as soluções apresentadas na literatura.\\
{\bf Palavras-chave}: Relatividade especial, simultaneidade, contração de lorentz,. 
\\
\\
\hspace{1cm}In this article, we present several apparent paradoxes of special relativity and their respective solutions. These paradoxes have appeared since the advent of relativity in 1905, and in fact they are never paradoxes. From a didactic point of view, paradoxes are an excellent learning tool. They lead the student to confront, and abandon, several central concepts of Galilean theory, such as simultaneity and rigidity of extended bodies. In particular, we review a new and simple solution to the twin paradox, recently found by one of the present authors \cite{TwinAlencar}.  It does not need to consider reference accelerators or light signals, which are the solutions presented in the literature.
\end{abstract}

\maketitle

\section{Introdução}
Desde o advento da Relatividade Especial (RE), em 1905, muitos paradoxos foram propostos. A expressão "paradoxos da relatividade" não é precisa, já que, de fato, ela é inteiramente consistente. Por "inteiramente consistente", queremos dizer que qualquer dos paradoxos pode ser resolvido somente com as transformações de Lorentz.  Esses paradoxos, de fato, emergem da tentativa de aplicar conceitos Galileanos em problemas intrinsecamente  relativísticos. Do ponto de vista didático, os paradoxos são uma excelente ferramenta de aprendizado. Veremos como cada paradoxo leva o estudante a confrontar, e abandonar, vários conceitos centrais da teoria Galileana, como a simultaneidade, a rigidez dos corpos extensos e a possibilidade de sinais se propagando com velocidade maior que a da luz. Nas próximas seções, apresentaremos as soluções a alguns desses paradoxos. Antes disso, faremos uma breve revisão do espaço-tempo Galileano, do espaço-tempo da RE, e da história de alguns desses paradoxos. 

\subsection{As Transformações de Galileu}
\begin{figure}[H]
\begin{center}
\includegraphics[scale=0.5]{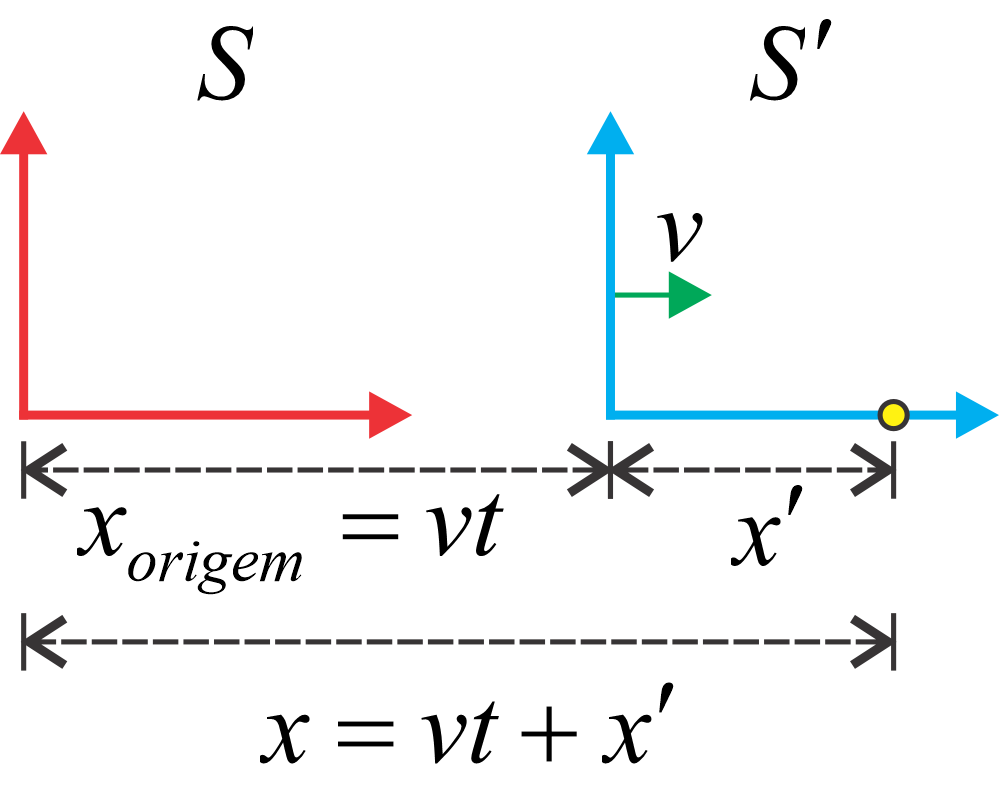}
\caption{Relação entre $S$ e $S'$}
\label{referenciaisgalileu}
\end{center}
\end{figure}

Considere referenciais $S$ e $S'$, com velocidade relativa $v$. Chamaremos de eventos as coordenadas $(t,x)$ e $(t',x')$, como descrito na figura \ref{referenciaisgalileu}.

A fim de encontrar as transformações de Galileu, que relacionam os eventos, como descritos por $S$ e $S'$,   precisamos da hipótese de que o tempo flui igualmente para os dois referenciais. Supomos, portanto, sem submeter ao crivo da experimentação, um tempo absoluto e \textit{universal}, ou seja, $t = t'$.
Após Galileu, Newton e todos os físicos usaram essa hipótese, que somente foi contestada por Einstein. Chamamos de hipótese pois, como veremos na próxima seção, isso só vale aproximadamente, para velocidades pequenas comparadas com a da luz.  
Na figura \ref{referenciaisgalileu}, vemos claramente que a posição de um ponto visto por $S$ é a posição do ponto visto por $S'$ somada à $x_{origem}$. Temos, então, a expressão     
 \begin{equation}\label{transformaçõesdegalileu}
        x= x' + x_{origem}=x' + v t\to x' = x -v t;\;t'=t
 \end{equation}
Essas são as \textit{transformações de Galileu}. Elas também podem ser escritas na forma de diferenças, relacionando  dois eventos
\begin{align}\label{transformaçõesdegalileudiferença}
&\Delta x =\Delta x' +v \Delta t';\;\Delta t'=\Delta t, &\\
\label{transformaçõesdegalileudiferençainversa}
&\Delta x' =\Delta x -v \Delta t;\;\Delta t'=\Delta t. &
 \end{align}
Acima, demos as transformações e suas inversas. 

Vejamos algumas consequências das transformações acima.  Primeiramente, temos uma que $\Delta t'=0\to \Delta t=0$. Portanto, se dois eventos são simultâneos em $S'$, também o serão em $S$. Outra consequência diz respeito ao comprimento de uma barra. Apesar de parecer simples, é importante definir de forma precisa o que é um comprimento. Consideremos uma barra em repouso em relação a um certo referencial inercial $S'$, com extremidades nas posições $x_2'$ e $x_1'$ respectivamente. Para que um observador nesse referencial meça o comprimento da barra, basta colocar uma régua sobre ela e marcar as extremidades. Dizemos que $\Delta x'=x_2'-x_1'=L'$. Para um referencial $S$, que observa $S'$, e portanto a barra, com velocidade $v$, é necessário que as posições dos extremos da barra sejam marcadas simultaneamente. Ou seja, o comprimento da barra em $S$ é definido por $\Delta t=0$. Com isso, a transformação (\ref{transformaçõesdegalileudiferençainversa}) implica $\Delta x'=\Delta x$ e  portanto $L'=L$.  Podemos dizer, portanto, que o espaço e o tempo Galileanos são absolutos. Na próxima seção veremos o caso relativístico

\subsection{As Transformações de Lorentz}

Podemos dizer a descoberta de Maxwell em 1861, de que a luz é uma onda eletromagnética,  inaugura a relatividade especial. O fato é que, a partir de 1861, se desenrolaram vários debates que culminaram, em 1905, na teoria da relatividade especial. Em sua formulação final, Einstein postula o princípio da relatividade de Galileu e que a velocidade da luz no vácuo é a mesma em todos os referenciais inerciais. Com isso, ele encontras as transformações de Lorentz e suas inversas\cite{RindlerLivro} (para uma demonstração acessível ao ensino médio, ver a Ref.  \cite{livro}, livro de um dos presentes autores.)  
\begin{align}
\label{tranfloren}
&\Delta x=\gamma\left(\Delta x' +v\Delta t' \right);\,\Delta t = \gamma\left(\Delta t' + \frac{v\Delta x'}{c^2}\right);\, \Delta z = \Delta z',&\\
\label{tranfloreninv}
&\Delta x'=\gamma\left(\Delta x -v\Delta t \right);\, \Delta t' = \gamma\left(\Delta t - \frac{v\Delta x}{c^2}\right);\, \Delta z = \Delta z',&
\end{align}
onde
\begin{equation}
    \gamma = \frac{1}{\sqrt{\left(1 - \frac{v^2}{c^2}\right)}}
\end{equation}
Uma importante propriedade das transformações acima é que, no limite $c/v\to0$, elas recuperam (\ref{transformaçõesdegalileudiferença}) e (\ref{transformaçõesdegalileudiferençainversa}) . Portanto, para baixas velocidades, e como esperado, podemos descrever o espaço-tempo com as transformações de Galileu. As transformações de Lorentz modificam profundamento nossa compreensão do espaço-tempo. Vejamos algumas dessas consequências.

\subsubsection{Simultaneidade, Dilatação do Tempo e o Paradoxo dos  Gêmeos}
Vejamos algumas consequências das transformações relativas ao tempo. Primeiramente vejamos a simultaneidade. No caso Galileano, vimos que $\Delta t'=0\to\Delta t=0$. Portanto, se dois eventos são simultâneos em $S'$, também o são em $S$. Considere agora o caso relativístico. Ao substituir $\Delta t'=0$ na Eq. (\ref{tranfloren}) obtemos 
\begin{equation}
\Delta t = \gamma \frac{v\Delta x'}{c^2}
\end{equation}
Portanto, diferente do caso Galileano, o fato de dois eventos serem simultâneos para $S'$ não implicam na simultaneidade para $S$.

Outra consequência das transformações Lorentz é a dilatação do tempo. Considere um relógio na origem de $S'$ e portanto $\Delta x'=0$. O tempo medido por esse referencial é chamado de tempo próprio $\Delta t'\equiv\Delta \tau$. A fim de comparar o tempo próprio, para os mesmos evento, com o tempo no referencial $S$, devemos usar as transformações de Lorentz. Da Eq. (\ref{tranfloren}),  podemos ver que $\Delta x'=0$ implica 
$$\Delta t=\gamma\Delta \tau.$$ 
Portanto, o tempo, do ponto de vista de $S$, se dilata. Isso vai de encontro com noção Galileana de tempo absoluto. 

A dilatação do tempo deu origem ao paradoxo dos gêmeos, certamente o mais famoso dentre todos. O ponto central desse paradoxo é que, do ponto de vista do referencial $S'$, o tempo de $S$ é que deveria se dilatar.  Imagine que Bob, o irmão gêmeo de Alice, parta da Terra para uma viagem intergalática. Para Alice, o tempo de Bob se dilata e portanto ele fica mais jovem. Do ponto de vista de Bob, o tempo de Alice dilata e ela é quem deveria ficar mais jovem. Ao retornar de viagem e compararem os relógios, quem será mais jovem?  Esse paradoxo apareceu poucos anos após a formulação da  Relatividade Especial(RE), em 1905. Ele tem sido continuamente discutido desde sua formulação padrão por Langevin, em 1911 \cite{Langevin}.  Uma solução completa, que não envolva referenciais acelerados, sinais emitidos entre os gêmeos e que utilize somente (\ref{tranfloren}), foi apresentada, bem recentemente, por um dos autores desse artigo \cite{TwinAlencar}. Veremos como a ausência de simultaneidade é suficiente para resolver esse paradoxo. Essa será a solução apresentada na seção \ref{gemeos}. 

Apesar de bastante contra intuitivas, a ausência de simultaneidade e a dilatação  do tempo são consequências diretas da Relatividade Especial. Como veremos nas próximas seções, a ausência de simultaneidade, particularmente, é central na resolução de muitos dos paradoxos. 

\subsubsection{Contração de Lorentz e os Paradoxos da  Contração}

Outra consequência das transformações (\ref{tranfloren}) é a contração de Lorentz.  Imagine uma barra horizontal, em repouso em relação ao referencial $S'$. Definimos o comprimento próprio da barra como o comprimento medido nesse referencial, ou seja, $L_0=\Delta x'$.  Para o referencial $S$, a barra tem velocidade $v$ e as extremidades da barra devem ser medidas simultaneamente, ou seja, $\Delta t=0$.  No caso Galileano, vimos que isso implicava que $L=L_0$. Todavia,  das transformações (\ref{tranfloreninv}), temos que $\Delta t=0$ implica
\begin{equation}
    \Delta x'=\gamma\Delta x.
\end{equation}
Como  $\Delta x=L$ , concluímos que $L=L_0/\gamma$ e portanto que $S$ observa a barra contraída. 

Um questionamento semelhante ao do paradoxo dos gêmeos pode ser feito aqui: do ponto de vista de $S'$, os comprimentos de $S$ é que devem se contrair.  É deveras curioso que, apesar da semelhança,  somente a partir dos anos 60 apareceram os primeiros paradoxos da contração. Em 1961, W. Rindler apresenta o primeiro deles \cite{Rindler}. Ele consiste em uma barra de comprimento próprio de 10 cm, que desliza sobre uma mesa horizontal, com uma velocidade tal que ${\gamma}=10$, enquanto se aproxima de uma fenda com mesmo comprimento próprio da barra. Ao levarmos em conta o referencial de repouso da fenda, a barra estará contraída, com um comprimento de 1cm e, evidentemente, cairá na fenda. Porém, ao tomarmos o referencial de repouso da barra, a fenda é vista com apenas 1cm, e a barra não cairá. Um pouco depois, a fim de evitar complicações do campo gravitacional, R. Shaw propõe uma versão simplificada desse paradoxo em 1962 \cite{Shaw}. O paradoxo é bem similar ao de Rindler, com a diferença de que a barra e a fenda possuem velocidades constantes, nas direções $x$ e $z$ respectivamente. Devido ao movimento ser em duas dimensões, as soluções para esses paradoxos envolvem, em geral, outras sutilezas da relatividade, como a mudança de ângulos. 

Como é comum na ciência, as versões mais simples dos paradoxos da contração apareceram posteriormente. Somente em 1963, em uma debate sobre o paradoxo das Naves de Bell, Dewan propôs uma versão unidimensional \cite{Dewan1963}. Por unidimensional, queremos dizer que o paradoxo envolve somente movimentos e comprimentos em uma direção.  Este ficou conhecido como o paradoxo do celeiro e da barra \cite{Dewan1963}. Ele consiste no seguinte: considere uma barra e um celeiro, ambos de comprimento próprio $L_0$. A barra se move com uma velocidade tal que, para o referencial do celeiro, ela terá comprimento $L_0/2$. Portanto, a barra cabe dentro do celeiro e a porta pode ser fechada após ela entrar. Todavia, do ponto de vista da barra, o celeiro deve ter comprimento $L_0/2$ e seria impossível fechar a porta. Após Dewan, outros paradoxos unidimensionais foram propostos. Dentre eles podemos citar o da fechadura e da chave. A versão unidimensional é a apresentada na maioria dos livros de relatividade. Muitas vezes com algumas variações, considerando um carro e uma garagem, ou guilhotinas cortantes.  

Consideraremos também paradoxos menos abordados nos livros textos: o das naves de Bell e, por fim, o das tesouras. Esse último, particularmente, é importante, pois envolve a noção de propagação de sinais com velocidades  maiores que a da luz.

\section{Uma nova solução para o paradoxo dos gêmeos}
\label{gemeos}
O paradoxo dos gêmeos, ou paradoxo do relógio, é um problema que apareceu poucos anos após a formulação da  Relatividade Especial (RE), em 1905. Ele tem sido continuamente discutido desde sua formulação padrão por Langevin, em 1911 \cite{Langevin}. O próprio Einstein achou uma solução qualitativa baseada na Relatividade Geral (RG), mas não encontrou um consenso. Nos anos 50, aconteceu um longo debate sobre o tema \cite{DINGLE}. Embora a resposta correta nunca esteve em questão, muitas soluções foram encontradas na literatura. Portanto, a questão de como resolver  o aparente paradoxo parece longe de um consenso. Para uma revisão ver a Ref. \cite{Shuler}. Muitas soluções acompanham Einstein e apelam para a aceleração do foguete \cite{Einstein}. Outras soluções se baseiam somente na RE e nossa solução é dessa classe. Em geral, essas soluções são de dois tipos: um envolve sinais de luz mandados pelo viajante para o gêmeo na Terra e o outro usa a relatividade da simultaneidade. Em ambos os casos diagramas do espaço-tempo são utilizados para resolver o problema. Estas soluções são apresentadas, por exemplo,  no famoso  livro do  Taylor e Wheeler \cite{Taylor}. As soluções que apelam para a aceleração parecem sugerir que a RG é necessária para resolver o paradoxo. Todavia, a RE é consistente por si só e alguma solução deveria existir sem apelar para a RG. Em certo momento, uma forma de evitar a aceleração do gêmeo foi encontrada \cite{Romer}. Todavia, o último autor não deu uma solução simples e analítica para o problema.  Uma solução que, de uma vez só, não envolva qualquer tipo de aceleração, sinais de luz ou diagramas do espaço tempo e seja baseada somente nas transformações de Lorentz foi apresentada recentemente por um dos presentes autores na Ref. \cite{TwinAlencar}. 

Vamos definir o problema. Temos dois gêmeos na Terra, Alice e Bob. Bob irá viajar para o planeta ``Ar" e voltará para a Terra. A distância própria, da Terra ao planeta Ar, é medida por Alice e dada por $L$. Os eventos relevantes são mostrados na  Fig. \ref{two} (Retirada da Ref. \cite{livro}). 
\begin{figure}[!h]
      \label{two}
        \centering
        \includegraphics[scale=0.7]{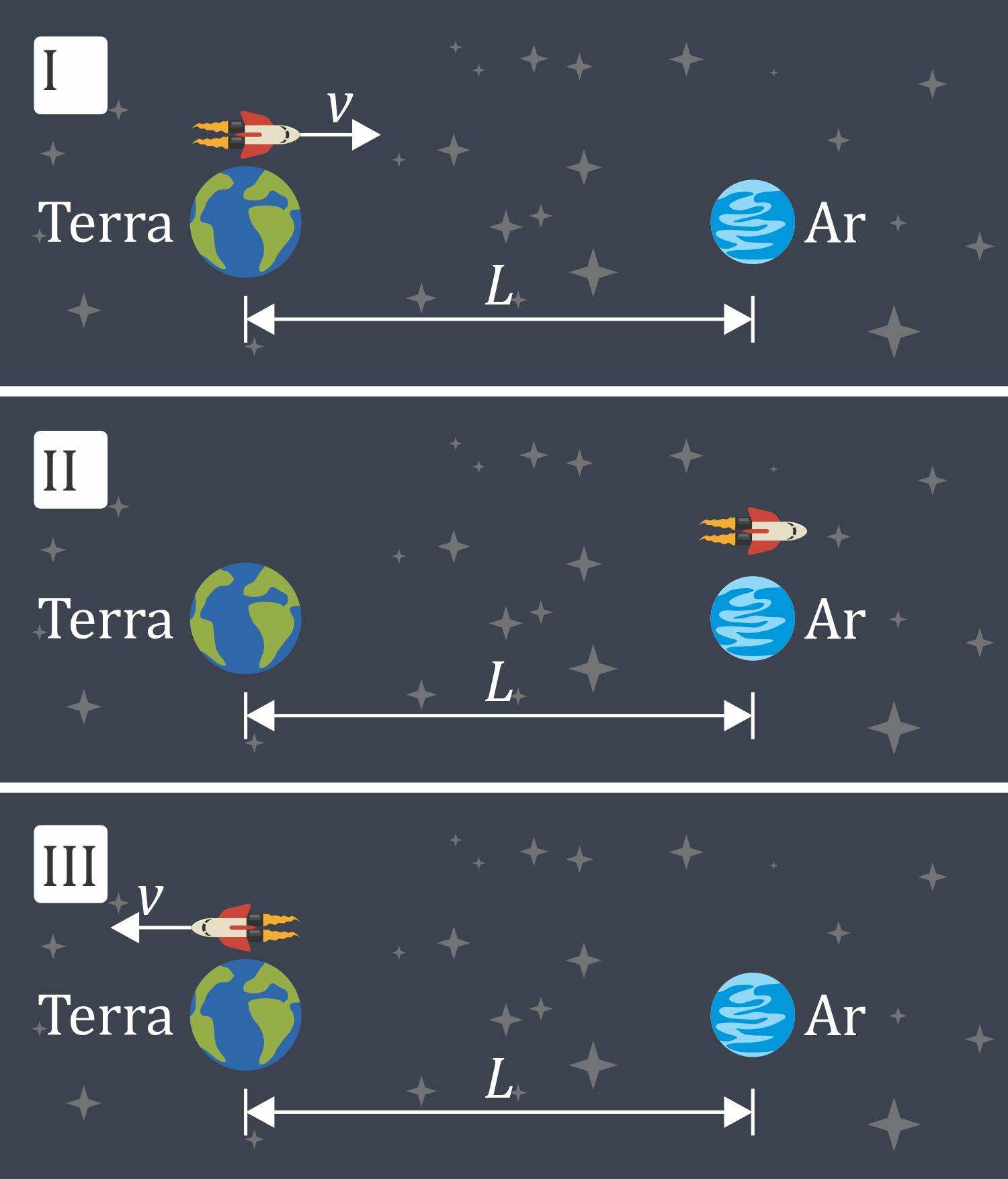}
        \caption{Os eventos relevantes, como descritos por Alice: $I$) Bob parte da Terra, $II$) Bob chega em Ar e $III$) Bob encontra Alice na Terra}
        \label{two}
 \end{figure}

Devemos ter muito cuidado pois, de acordo com referenciais diferentes, cada relógio marca tempos diferentes. Portanto, sempre que falarmos em medidas de tempo, devemos indicar o relógio e o observador. Vejamos primeiramente os eventos segundo Alice. O tempo de viagem de Bob, marcado no relógio de Alice e de acordo com ela, é dado por 
\begin{equation}\label{relogioAlice}
 \Delta T_A=2L/v,   
\end{equation}
onde  $v$ é a velocidade do foguete. Todavia,  para Alice, o relógio de Bob tem velocidade $v$ e, devido à dilatação do tempo, marcará 
\begin{equation}
    \Delta T_B=\frac{\Delta T_A}{\gamma}=\frac{2L}{v\gamma}
\end{equation}
onde $\gamma$ é o fator de Lorentz. Portanto, Alice afirmará que Bob estará  mais jovem quando voltar. E o que Bob acha? Do ponto de vista dele, devido à contração de Lorentz, a distância entre a Terra e o planeta Ar é dada por  $L/\gamma$.  Portanto o relógio de Bob, segundo ele, marcará um tempo de viagem dado por  
\begin{equation}
\label{tempobobbob}
\Delta T'_B=\frac{2L}{v\gamma}.
\end{equation}

Ambos concordam, portanto, com o tempo de viagem de Bob. Todavia, para Bob, o relógio de Alice é que tem velocidade $v$ e deve dilatar pelo fator 
\begin{equation}
\label{tempoaliceparabob}
 \Delta T'_A=\frac{\Delta T'_B}{\gamma}=\frac{2L}{v\gamma^2}.   
\end{equation}

Quando eles se encontrarem, o relógio de Alice irá mostrar $2L/v$ (Eq.(\ref{relogioAlice})) ou $2L/(v\gamma^2)$ (Eq.(\ref{tempoaliceparabob}))? Este é o paradoxo. Alice diz que Bob voltará mais jovem devido à sua velocidade. De acordo com Bob, é Alice quem tem velocidade e portanto será a mais jovem. O ponto é: quando eles encontrarem e compararem o relógio, quem estará correto?

\begin{figure}[!h]
        \centering
        \includegraphics[scale=0.7]{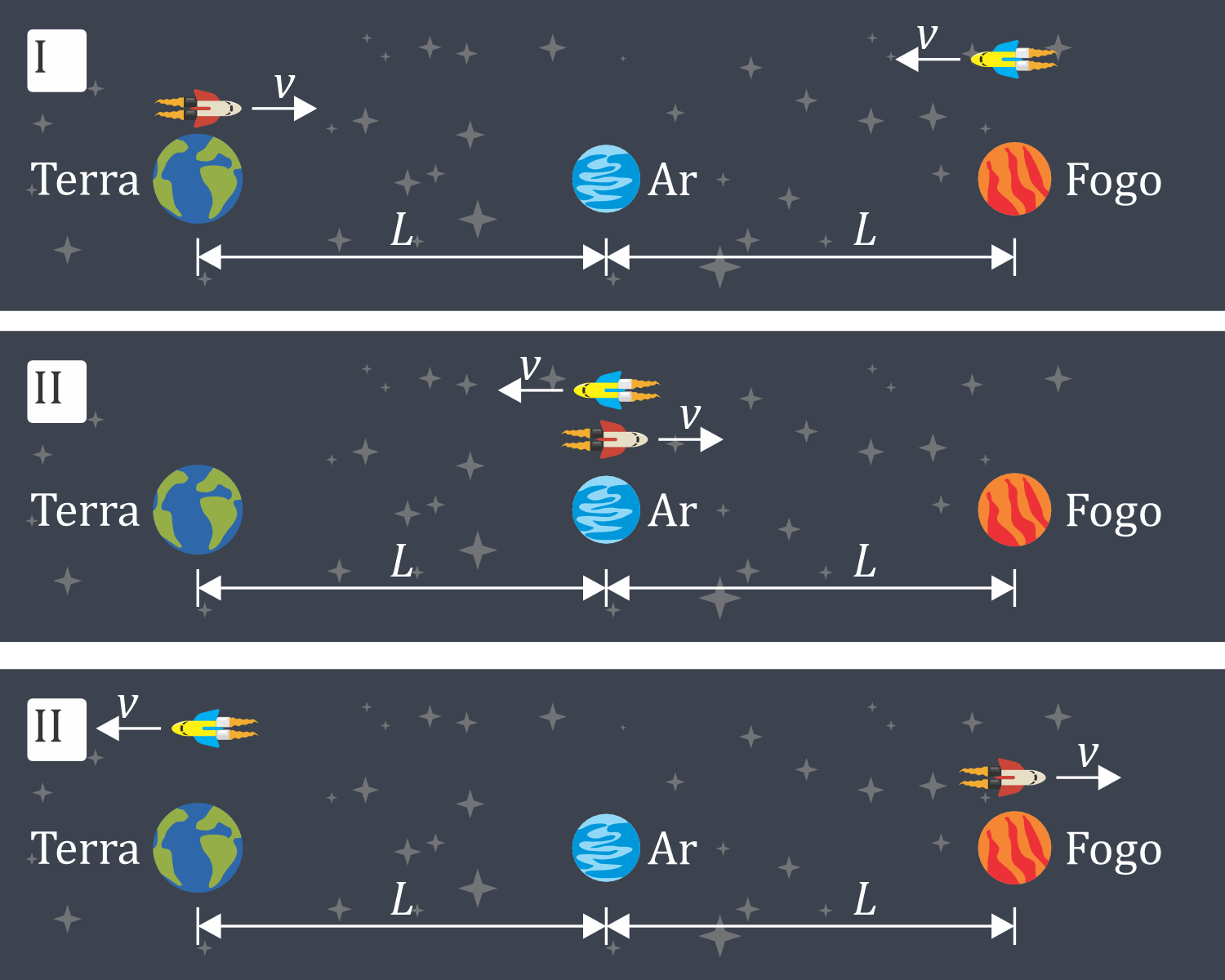}
        \caption{Os eventos relevantes, como descritos por Alice: $I$) Bob e Marcos partem da Terra e ``Fogo'', $II$) Bob e Marcos chegam em ``Ar" e $III$) Marcos encontra Alice na Terra e Marcos chega em Fogo.}
           \label{three}
 \end{figure}

 Como dito antes, nós evitaremos aceleração e utilizaremos somente as transformações de Lorentz. A forma de fazer isso foi apontada por Romer \cite{Romer}. Nós utilizaremos uma configuração um pouco diferente.  A fim de evitar referenciais acelerados, consideraremos um terceiro gêmeo, Marcos, viajando de um terceiro planeta, chamado ``Fogo''. Este também está em repouso em relação à Terra. Os planetas Ar e Fogo estão a uma distância   $L$ e  $2L$ da Terra. Os eventos relevantes estão na Fig. \ref{three} (Retirada da Ref. \cite{livro}).

A fim de sermos mais precisos, usaremos a notação $I_T$ para o evento $I$ na Terra, no referencial da Alice. Uma linha, $I'_T$, significará o mesmo evento no referencial do gêmeo Marcos. Para Alice, as partidas de Bob e Marcos são simultâneas, com velocidades $v$ e $-v$ respectivamente. Portanto, os eventos  $I_T$ e $I_F$ são simultâneos para ela. Desta forma, Marcos chegará na Terra ao mesmo tempo que Bob chega em Fogo e os eventos $III_T$ and $III_F$ também são simultâneos para ela. Claro, Alice dirá que Bob e Marcos tem a mesma idade. Isso se deve ao fato de que as velocidades relativas, em relação a ela, são as mesmas. Isso simplifica bastante o problema, já que Marcos não tem aceleração e os relógios que serão comparados são os de Alice e Marcos, no evento $III_T$. Claro, poderíamos imaginar outra situação, em que Bob para em Ar e volta para a Terra. Todavia, isso é irrelevante, já que, para Alice, ambos devem ser mais jovens e ter a mesma idade. Aí reside a solução, que usa, portanto, somente referenciais inerciais. Vamos focar, então, em Alice e Marcos. 

Como dito acima, para Alice, os eventos $I_T$ e $I_F$ são simultâneos e portanto $\Delta T=0$. Todavia, para Marcos, a Terra e Fogo tem velocidades $-v$. Para ele, temos de acordo com as transformações de Lorentz temos
\begin{equation}\label{simultaneidade}
\Delta T'=\gamma(\Delta T+\frac{v}{c^2}\Delta x)=\gamma\frac{2Lv}{c^2}.
\end{equation}
Na segunda igualdade usamos $\Delta T=0$ e $\Delta x=2L$. Portanto, de acordo com Marcos,  o início da marcação no seu relógio (evento $T'_{I'_F}$), e no de Alice (evento $T'_{I'_T}$), não são simultâneos. A diferença de início, devido à ausência de simultaneidade, é dada pela equação acima. De forma mais detalhada teremos
\begin{equation}\label{eq01}
T_{I'_T}-T_{I'_F}=\gamma\frac{2Lv}{c^2}.
\end{equation}
Esta é a diferença de tempo no relógio de Marcos. Lembre agora que, segundo ele, o tempo no relógio de Alice se dilata por um fator $1/\gamma$. Portanto, quando Marcos parte, ele diz que o relógio de Alice estará marcando $2 Lv/c^2$. Para descobrir quanto estará marcado o relógio de Alice, na chegada de Marcos, devemos agora adicionar o tempo de viajem. Como dito antes, do ponto de vista de Bob, esse tempo é dilatado e dado pela Eq.(\ref{tempoaliceparabob}). Ao somar os tempos teremos  
\begin{equation}\label{eq02}
\frac{2 Lv}{c^2}+\frac{2L}{v\gamma^2}=\frac{2L}{v}.
\end{equation}

 Quando Marcos encontra Alice, o relógio dela estará mostrando $2L/v$. Já o relógio de Bob, segundo ele, marcará somente o tempo de viagem, dado por (\ref{tempobobbob}). Podemos afirmar, então, que:
 \\
 \\
 \textbf{Do ponto de vista dos dois referenciais, ambos concordas que o relógio Alice estará marcando $2L/v$ e o de Marcos $2L/(v\gamma)$.}
 \\
 \\
 Nós concluímos que, apesar de Marcos ver o tempo do relógio de Alice passando mais lentamente, todos concordam com o fato de que Marcos, e portanto Bob, são mais jovens. Isso se deve à ausência de simultaneidade e, portanto, para Marcos o relógio de Alice começou a marcar o tempo antes. Para ele, ela estará mais velha. É deveras interessante que, para Marcos, a ausência de simultaneidade compense exatamente a dilatação do tempo de Alice, de tal forma que todos concordem que ela estará mais velha. Paradoxo belamente resolvido!

Finalmente, devemos assinalar que Romer alcançou um passo importante ao evitar aceleração \cite{Romer}.  Todavia, neste artigo, ele somente sugere que a simultaneidade  \textbf{deveria} resolver o problema, mas não apresenta o cálculo exato, nem a explicação detalhada dos eventos. Essa atitude está presente praticamente todas a soluções encontradas na literatura. Os autores nunca mostram que a simultaneidade é \textbf{exatamente} suficiente para resolver o paradoxo. Isso foi sanado na Ref. \cite{TwinAlencar}, por um dos presentes autores, e  aqui apresentamos a solução.  

\section{Paradoxos da Contração em uma Dimensão}

Como dissemos na introdução, devemos primeiramente considerar os paradoxos da contração unidimensionais, por serem de mais fácil resolução. Apresentaremos a solução para três paradoxos da contração. Dois deles são protótipos para a resolução de muitos outros paradoxos. A versão apresentada aqui foi dada na Ref. \cite{livro}, de um dos presentes autores. Diferente do caso do paradoxo dos gêmeos, muitos livros e artigos apresentam boas e detalhadas soluções\cite{Kneubil}. O terceiro, o da chave e fechadura, pode ser considerado uma aplicação dos anteriores.  

\subsection{Paradoxo da madeireira relativística}
\label{madeireira}
Imagine que uma madeira, de comprimento próprio $L_0$, esteja em uma esteira relativística. A madeira passará entre duas guilhotinas, com distância própria também $L_0$. Alice opera a guilhotina e Bob, aventureiro, sobe na esteira para observar o que ocorre. O paradoxo consiste no seguinte: De acordo com Alice, a madeira terá velocidade $v$ e, devido à contração de Lorentz, comprimento $L_0/\gamma$. Portanto, ela será menor que a distância $L_0$ entre as guilhotinas. Alice diz que a madeira passa pelas guilhotinas sem ser cortada. Já para Bob, é a guilhotina que tem velocidade $-v$ e, portanto, comprimento $L_0/\gamma$. Para ele, a madeira tem comprimento $L_0$ e, sendo maior que  a distância entre as guilhotinas, será cortada. A madeira será cortada ou não?

\begin{figure}[h]
    \centering
    \includegraphics[width=6cm]{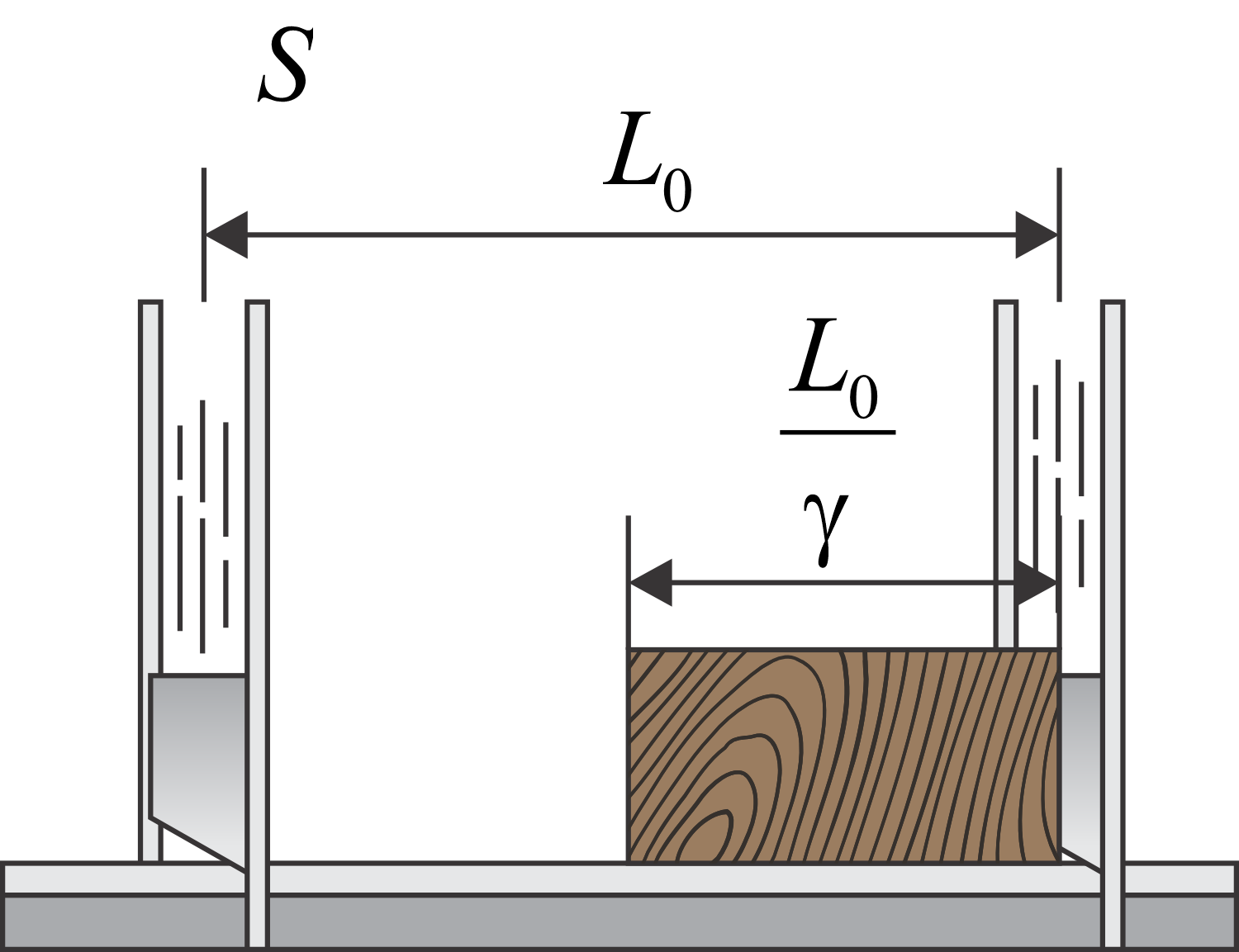}
    \caption[Figura 2]{Configuração do sistema em S}
    \label{madeiraalice}
\end{figure}

A solução para esse paradoxo, assim como no caso do paradoxo dos gêmeos, reside na simultaneidade. Veremos que, se para um referencial a madeira é cortado, para o outro ela também deve ser cortada. Vejamos a situação em que, para Alice, ela não seja cortada.  Imagine que, para Alice, a guilhotina da direita baixe no exato momento em que a extremidade direita da madeira chegue nela. Para Alice, as guilhotinas baixam simultaneamente, e, devido à contração da madeira, a guilhotina da esquerda não vai cortá-la. Isso é mostrado na Fig. \ref{madeiraalice} (Retirada da Ref. \cite{livro})

\begin{figure}[h]
    \centering
    \includegraphics[width=8cm]{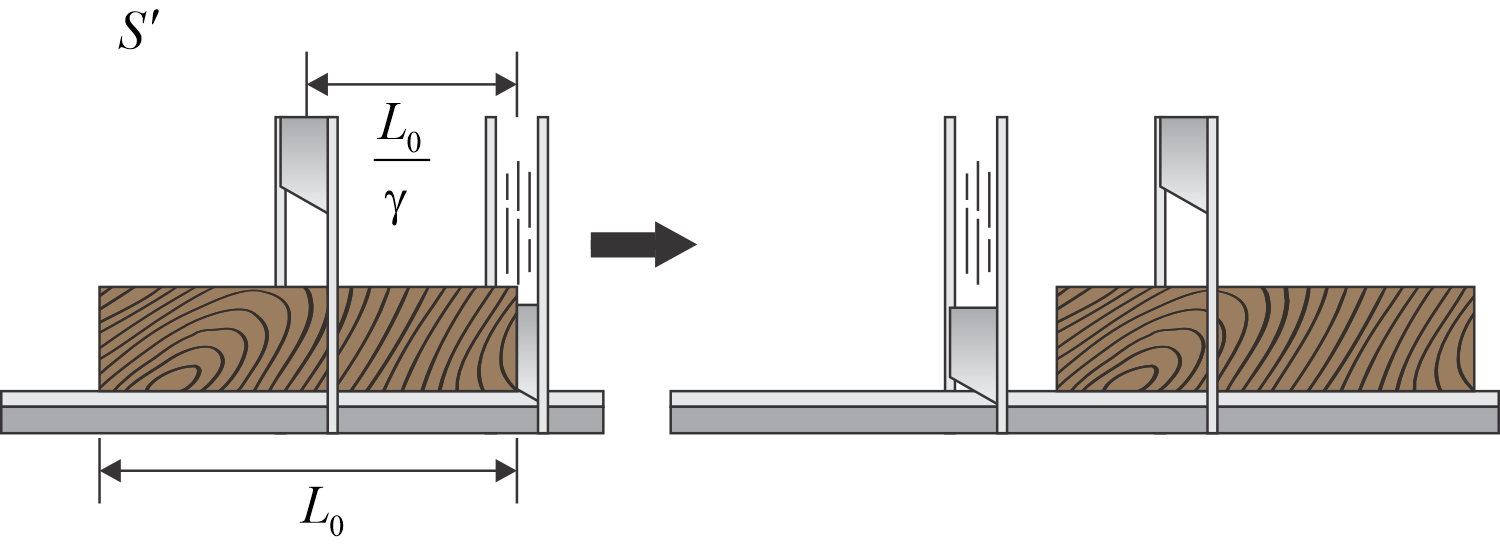}
    \caption[Figura 3]{Configuração do sistema em S'}
    \label{madeirabob}
\end{figure}

Vejamos o que ocorre para Bob. Segundo ele, a distância entre as guilhotinas é $L_0/\gamma$ e a madeira tem comprimento $L_0$. Todavia, para ele, as guilhotinas não baixam simultaneamente. O problema é bastante semelhante ao tratado na seção anterior. Vimos que a diferença de tempo é dada pela Eq.(\ref{simultaneidade}), ou $\Delta T=(\gamma L_0 v)/c^2$. Nesse caso utilizamos $\Delta x=L_0$. Ou seja, a guilhotina da direita baixa antes da guilhotina da esquerda. No caso dos gêmeos, foi o relógio de Alice que começou a marcar o tempo antes. A situação como vista por Bob, é mostrada na Fig. \ref{madeirabob} (Retirada da Ref. \cite{livro})

A pergunta é: até que a guilhotina da esquerda baixe, teremos tempos suficiente para que a madeira passe sem ser cortada? Vejamos. Claro que, tanto para Bob quanto para Alice, a guilhotina da direita baixa no mesmo momento em que  extremidade direita da madeira chega lá. São eventos que ocorrem no mesmo ponto. Portanto, no momento em que a extremidade direita baixa, o comprimento  da madeira, que estará para ``fora'' da guilhotina, é dada por 

\begin{equation}\label{excesso}
d_1=L_0-L_0/\gamma.
\end{equation}

Todavia, vejamos a distância $d_2$ que a madeira anda até que a guilhotina da esquerda baixe. Isso é um problema muito simples: temos a velocidade da madeira e o tempo total. Com isso, teremos

\begin{equation}
d_2=v\times \Delta T=\frac{\gamma L_0 v^2}{c^2}=\gamma L_0-\frac{L_0}{\gamma }\to d_2-d_1= L_0(\gamma -1).    
\end{equation}

Como $\gamma>1$, $d_2>d_1$ e a madeira passa sem ser cortada. Maior quanto? Tanto quanto se queira, basta que consideremos uma velocidade $v$ cada vez mais próxima de $c$.

É possível ainda considerar que, para Bob, as guilhotinas baixem simultaneamente e portanto a madeira é cortada. Nesse caso, é possível mostrar que, apesar de ver a madeira menor que a distância entre as guilhotinas, Alice concorda que ela será cortada. Esse é um bom exercício. 

\subsection{A guilhotina emperrada}
\label{emperrada}
Consideraremos o caso em que a guilhotina da direita emperra. Esse seria o equivalente ao paradoxo original, do celeiro e da barra, proposto na Ref. \cite{Dewan1963}. Também é equivalente ao do  carro e da garagem e a muitos outros propostos em livros textos. Para Alice, a madeira continua sem ser cortada. Todavia, para Bob, se a guilhotina emperrar, ela deve ``arrastar'' a madeira para a esquerda. Quando a segunda guilhotina baixar, ela irá cortar a madeira? 

A solução para esse paradoxo reside na não rigidez de corpos extensos na relatividade. No nosso caso, isso se deve ao fato de que nada pode viajar mais rápido que a velocidade da luz. Portanto, até que a informação de que a extremidade da esquerda dever ser ``arrastada", a da direita já caminhou uma distância $d_2$. Podemos imaginar o corpo formado por varias camadas de átomos. Ao se chocar com a guilhotina, a primeira camada empurra a segunda, na sequencia a segunda camada empurra a terceira,  e assim por diante, até chegar à extremidade esquerda. 

Mas o tempo será suficiente, do ponto de vista de Bob, para que a madeira fique entre as guilhotinas e não seja cortada? Podemos considerar o caso limite, em que, ao bater na extremidade direita, a informação se propague para a extremidade esquerda com a velocidade da luz. Portanto, essa extremidade só é arrastada depois de um tempo $T_1=L_0/c$ e, portanto, percorre a distância 
\begin{equation}
    d_3=L_0\frac{v}{c} 
\end{equation}
Agora podemos avaliar se essa distância é maior que o excesso de comprimento da madeira, dado por $d_1$ na Eq.(\ref{excesso}). Temos
\begin{equation}
d_2-d_1= \frac{L_0}{\gamma}\left(1-\frac{1}{(1+\frac{v}{c})\gamma}\right)>0.    
\end{equation}

A distância percorrida pela extremidade esquerda, até que a guilhotina baixe, é maior que o excesso de tamanho da madeira. Do ponto de vista de Bob, portanto, ela também não é cortada. O paradoxo acima levanta uma questão importante na relatividade, que é a modificação da noção de corpo rígido na relatividade especial. 

\subsection{Paradoxo da chave e fechadura}
\label{chave}
\begin{figure}
    \centering
    \begin{subfigure}{0.3\textwidth}
        \includegraphics[width=\textwidth]{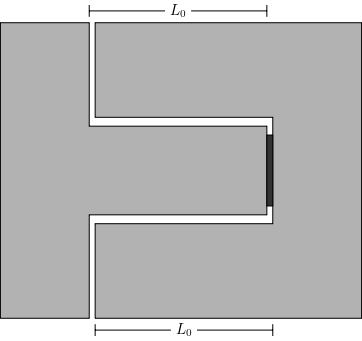}
        \caption{Sistema no referencial de repouso de ambos}
        \label{chave1}
    \end{subfigure}
    
    \begin{subfigure}{0.3\textwidth}
        \includegraphics[width=\textwidth]{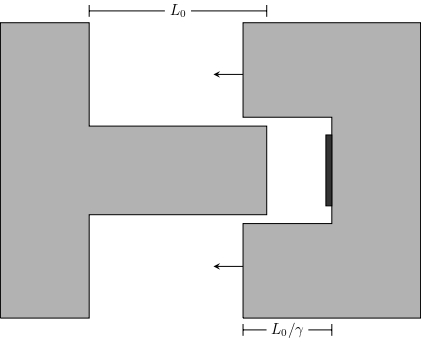}
        \caption{Sistema no referencial de repouso da chave}
        \label{chave2}
    \end{subfigure}
    
    \begin{subfigure}{0.3\textwidth}
        \includegraphics[width=\textwidth]{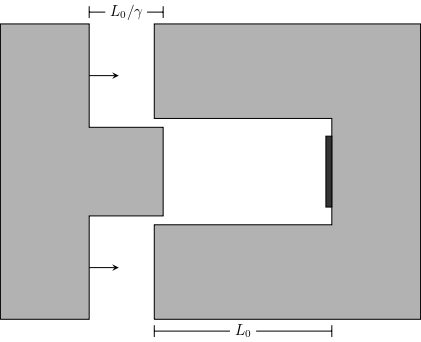}
        \caption{Sistema no referencial de repouso da fechadura}
        \label{chave3}
    \end{subfigure}
    \caption{Configuração do sistema chave e fechadura}
\end{figure}

Esse paradoxo é mais uma versão do paradoxo celeiro e da barra. De autoria desconhecida, ganhou maior projeção ao ser apresentado como exercício no famoso livro de E. F. Taylor e J. A. Wheeler, em 1992 \cite{Taylor}. Mais recentemente, ele é discutido em 2007, por E. Pierce \cite{Pierce}. Ele consiste em uma chave em \textbf{T} e em uma fechadura em \textbf{U}, com um botão na extremidade interna. A porta se abre apenas se a chave tocar o botão. Em repouso, ambas, a chave e a fechadura, possuem comprimentos próprios $L_0$ e portanto a chave tem o tamanho exato para apertar o botão, como na Fig. \ref{chave1}. Portanto, a porta se abre. Considere agora que exista uma velocidade relativa entre a chave e a fechadura. No referencial de repouso da chave, a fechadura se contrai, e o botão ainda é alcançado, como no caso da Fig. \ref{chave2}. Todavia, no referencial em que a fechadura está em repouso, a chave se contrai e o botão não é pressionado, como visto no caso da Fig. \ref{chave3}. Nisso consiste o paradoxo. 

Sua solução é bastante semelhante à da guilhotina emperrada, na seção \ref{emperrada}. Baseia-se, portanto, na velocidade finita de propagação da informação ao longo da chave. Considere a situação da Fig. \ref{chave3}. No momento em que a haste horizontal da chave atingir a borda da fechadura, a ponta continuará se movendo brevemente até que a informação a alcance. A pergunta, mais uma vez, é: esse tempo será suficiente para a extremidade da direita pressionar o botão? Vejamos. 

A chave, de comprimento próprio $L_0$, viaja a uma velocidade $v$, no referencial $S$. Sua extremidade traseira é subitamente freada até o repouso. A distância entre a chave e o botão portanto será exatamente igual a (\ref{excesso}). Assumindo que a informação viaje na velocidade da luz ao longo da chave, o tempo para que ela atinja a extremidade dianteira da barra é:
\begin{equation}
    \label{eq:12}
    ct = \frac{L_0}{\gamma} + vt \Rightarrow t = \frac{L_0}{\gamma(c-v)}
\end{equation}
Nesse intervalo, a extremidade dianteira viajou uma distância dada por:
\begin{equation}
    d_2 = vt = \frac{vL_0}{\gamma(c-v)}
\end{equation}
Portanto, se $d_2-d_1>0$, nosso paradoxo está resolvido. Vejamos.
\begin{equation}
    d_2-d_1 = \frac{vL_0}{\gamma(c-v)} + \frac{L_0}{\gamma}-L_0=\frac{L_0}{\gamma(1-\frac{v}{c})} -L_0
\end{equation}
ou
\begin{equation}
    d_2-d_1 = L_0\left[\gamma(1+\frac{v}{c}) -1\right]>0
\end{equation}

Portanto, até que a informação de parar chegue na extremidade direita, ela já tocou o botão e a porta se abriu. O paradoxo está, portanto, resolvido. 

\section{Paradoxos da contração em duas dimensões}
Nessa seção, abordaremos os paradoxos que envolvem movimento em duas dimensões. Eles possuem algumas sutilizas que não aparecem no caso unidimensional. Primeiramente iremos considerar o caso que não envolve aceleração, como apresentado em 1962 por Shaw \cite{Shaw}. Só posteriormente trataremos do artigo seminal de Rindler, que envolve aceleração, e é o primeiro dos paradoxos da contração \cite{Rindler}. 
    
    \subsection{Barra e Fenda sem efeito da gravidade}
    \label{barraefenda}
\begin{figure}[H]
	\centering
	\includegraphics[width=1\linewidth]{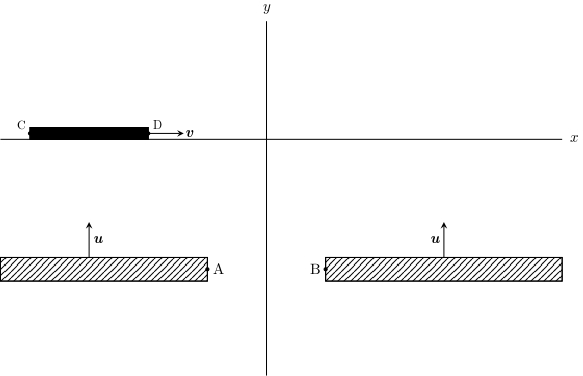}
	\caption[Figura 5]{Configuração do sistema no referencial inicial}
	\label{fig:barra_e_fenda1}
\end{figure}

	Considere uma barra, de comprimento próprio $L_0$ e velocidade \textit{\textbf{v}}, que se move ao longo de seu comprimento, na direção $x$ do referencial $S'$. Para o mesmo referencial,  uma plataforma com uma fenda de comprimento próprio $L_{0}$ e paralela à barra, realiza um movimento vertical para cima com velocidade \textbf{\textit{u}}. Essa situação é ilustrada na Fig. \ref{fig:barra_e_fenda1}. Como a barra possui movimento na direção do seu comprimento, ela terá comprimento $L_0/\gamma$. Portanto, a barra deve passar pela fenda. Imagine agora o referencial em repouso com a barra. Para ele, a fenda terá comprimento $L_0/\gamma$ e, portanto, a barra não deveria passar.  
	
	A solução para esse paradoxo envolve uma nova sutileza da relatividade: a modificação da inclinação de uma barra em movimento. Como ilustração imagine que, para o referencial $S'$, uma barra  possui um ângulo $\tan \theta'=L_0{_y}/L_0{_x}$ com o eixo $x$. Para o referencial $S$ a barra se move para a direita, e portanto a componente horizontal se contrai $L{_x}=L_0{_x}/\gamma$, já a componente vertical não e $L{_y}=L_0{_y}$. Portanto, para o referencial $S$ o ângulo se modifica para $\tan \theta=\gamma\tan \theta'$. Note que $\theta'=0\to\theta=0$ e que essa sutileza não aparece nos casos unidimensionais. 
	
	Existe alguma situação em que, mesmo tendo $\theta'=0$ para $S'$, teremos $\theta\neq 0$ para $S$? A resposta é sim, e aí reside a solução para nosso paradoxo. Isso ocorre na situação descrita na Fig. \ref{fig:barra_e_fenda1}. Do ponto de vista de $S'$,  a plataforma terá uma inclinação e isso propiciará que a barra passe pela fenda. Vejamos, matematicamente, como isso ocorre. Lembre que, para o referencial $S'$, a plataforma e a fenda devem ser medidos de forma simultânea. Considerando as extremidades $A$ e $B$ da fenda isso significa, para esses eventos,  $\Delta t'=0$. Essa foi a mesma estratégia usada, na introdução,  para obter a contração de uma barra que se move horizontalmente. Substituindo isso nas transformações de Lorentz (\ref{tranfloren}) obtemos
	\begin{equation}
	 \Delta x = \gamma \Delta x'\to  \Delta x'= L_0/\gamma  
	\end{equation}
	Na segunda igualdade acima, usamos o fato de que o comprimento próprio é medido em $S$, logo $\Delta x=L_0$. Portanto, temos a contração usual, devido ao movimento a longo de um comprimento. 
	
	Vejamos agora o que ocorre com as componentes $y_A,y_B$ e $y'_A,y'_B$. Das transformações de Lorentz vimos que $y=y'$, e portanto temos 
	\begin{equation}
	   y'_A=y_A=ut_A, \;y'_B=y_B=ut_B \to \Delta y' =\Delta y=u \Delta t
	\end{equation}
	Em um primeiro momento, somos levados a pensar que 
 \begin{equation}
     \Delta t =t_B-t_A=0\to\Delta y' =\Delta y=0
 \end{equation}
	Portanto, não teríamos inclinação no referencial $S'$.  Todavia, $\Delta t=0$ é a descrição da plataforma pelo referencial $S$. A fim de $S'$ medir a plataforma simultaneamente, tivemos que impor $\Delta t'=t'_B-t'_A=0$. Das transformações de Lorentz (\ref{tranfloren}) isso implica, necessariamente, que
	\begin{equation}
	 \Delta t'=\gamma \left(\Delta t-\frac{v}{c^2}\Delta x\right)\to \Delta t=\frac{v L_0}{c^2}.
	\end{equation}
	Na última igualdade usamos mais uma vez que $\Delta x=L_0$. Com isso, obtemos nosso resultado final, dado por 
	\begin{equation} \label{projecoes}
	\Delta y'=u\Delta t = \frac{uvL_0}{c^2};\; \Delta x'= L_0/\gamma.
	\end{equation}
	Analisando as equações acima, é possível perceber que a orientação da barra no referencial $S'$  difere da orientação observada em $S$, devido à ausência da simultaneidade. Logo, podemos calcular a inclinação da barra em $S'$.
	\begin{equation}\label{eq:inclinação}
	 \tan\alpha' = \frac{\Delta y'}{\Delta x'}  = \frac{uv\gamma}{c^2}	 
 	\end{equation}
 	
\begin{figure}[H]
		\centering
		\includegraphics[width=1\linewidth]{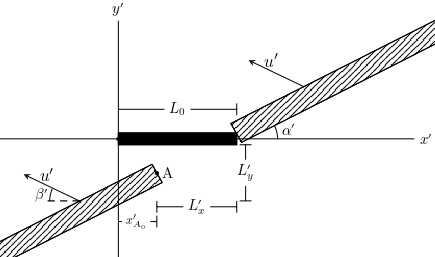}
		\caption[Figura 6]{Configuração do sistema no referencial de repouso da barra}
		\label{fig:barra_e_fenda2}
	\end{figure}

 	Outro detalhe importante é a mudança na direção da velocidade $\textbf{u'}$. Suas componentes são relacionadas, utilizando a transformação de velocidades relativística, por
		\begin{equation}\label{velocidadesfenda}
		    u'_x = -v;\;	u'_y = \frac{u}{\gamma}
		\end{equation}
 	Portanto, obtemos 
 	\begin{equation}
		 \tan \beta'=-\frac{u'_y}{u'_x}=\frac{u}{v\gamma}
	\end{equation}
 	A Fig. \ref{fig:barra_e_fenda2} esquematiza o sistema visto pelo referencial $S'$, de modo que a inclinação é dada por (\ref{eq:inclinação}).

\begin{figure}[H]
		\centering
		\includegraphics[width=0.9\linewidth]{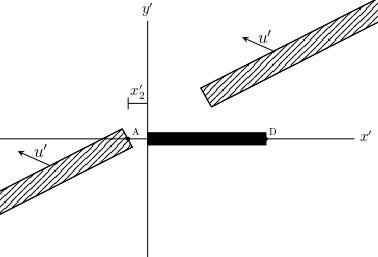}
		\caption[Figura 7]{Configuração do sistema após um dado intervalo de tempo}
		\label{fig:barra_e_fenda3}
	\end{figure}

		Com os resultados acima podemos analisar, de acordo com $S'$,  o movimento da fenda. Consideremos que, em $t'=0$, a extremidade $D$ da barra passe bastante próximo da extremidade $B$ da fenda, como na Fig. \ref{fig:barra_e_fenda2}. Por simplicidade, coloquemos a origem do eixo $x'$ na extremidade esquerda da barra, o ponto $A$. Depois de um tempo, que chamaremos $t'_2$, a extremidade $A$ da fenda cruza o eixo $x'$ na posição $x'_2$. Essa configuração é descrita na Fig. \ref{fig:barra_e_fenda3}.
    
    Fica claro, portanto, que  a barra passa pela fenda se $x'_2<0$. Para obter isso, primeiramente precisamos do tempo necessário para a extremidade $A$ cruzar o eixo $x'$. Utilizando as  Eqs.(\ref{projecoes}) e  (\ref{velocidadesfenda}), obtemos facilmente 
    \begin{equation} \label{t2}
       t_2' = \frac{\Delta y'}{u'_y} = \frac{L_0 v \gamma}{c^2} 
    \end{equation}
    Para determinar a posição $x'_2$, precisamos da posição inicial do ponto $A$. Na configuração acima, e utilizando (\ref{projecoes}),  vemos que ela é dada por 
    \begin{equation} \label{x20}
    x'_{A_0}=L_0-\frac{L_0}{\gamma}.   
    \end{equation}
    Das Eqs.(\ref{t2}) e (\ref{x20}) obtemos, então,  qual a posição $x'_2$
	\begin{equation}
      x'_2 =x'_{A_0}+u'_x t_2' = L_0-\frac{L_0}{\gamma}-\frac{L_0 v^2 \gamma}{c^2} =L_0(1-\gamma)
    \end{equation}
	Como $\gamma \geq 1$ temos  $x'_2<0$, e sempre  a barra passará pela fenda. 
	
\begin{figure}[H]
	\centering
	\includegraphics[width=0.9\linewidth]{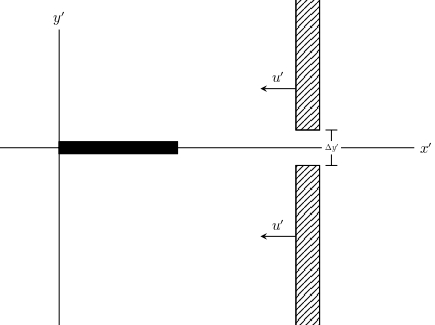}
	\caption[Figura 6]{Configuração do sistema quando o fator de Lorentz tende ao infinito}
	\label{fig:barra_e_fenda4}
\end{figure}

	Um caso interessante ocorre quando $v\to c$, e, portanto, $\gamma\to \infty$. Nesse caso, a Eq.(\ref{eq:inclinação}) implica que $\tan \alpha'$ tenderá ao infinito. Isso resulta em uma rotação de $\pi/2$ e a plataforma fica na vertical. Já a Eq.(\ref{projecoes}) nos da que $\Delta x'=0$ e $\Delta y'=uL_0/c$. Desse modo, a plataforma fica vertical e a abertura da fenda será $uL_0/c$. Essa situação é descrita na Fig. \ref{fig:barra_e_fenda4}.

\subsection{Barra e Fenda com efeito da gravidade}
\label{barraefendagrav}

Finalmente, vamos tratar do artigo seminal, de 1961, de Rindler \cite{Rindler}. Devido à sua importância, discutiremos duas soluções. A dada pelo próprio Rindler, no artigo em que propõe o paradoxo, e uma versão simplificada  discutida por Leo Sartori \cite{Sartori}. 

\subsubsection{Solução de Rindler}

Nesse caso, originalmente proposto pro Rindler, temos uma barra horizontal deslizando, com velocidade \textit{v},  sobre uma plataforma que possui uma fenda. A barra e a fenda possuem comprimento próprio $L_0$. Seja \textit{S} o referencial de um observador parado em relação à plataforma e \textit{S'} o referencial que viaja horizontalmente junto à barra. Além disso, para evitar que a barra sofra rotação durante a queda, considera-se que a fenda está coberta por um alçapão que abre imediatamente quando a barra está inteiramente sobre ele. Dessa forma, todos os pontos da barra caem igualmente, e ela permanece horizontal. Assume-se ainda que o campo gravitacional pode ser tratado como uniforme.
\begin{figure}
    \centering
    \includegraphics[scale=0.77]{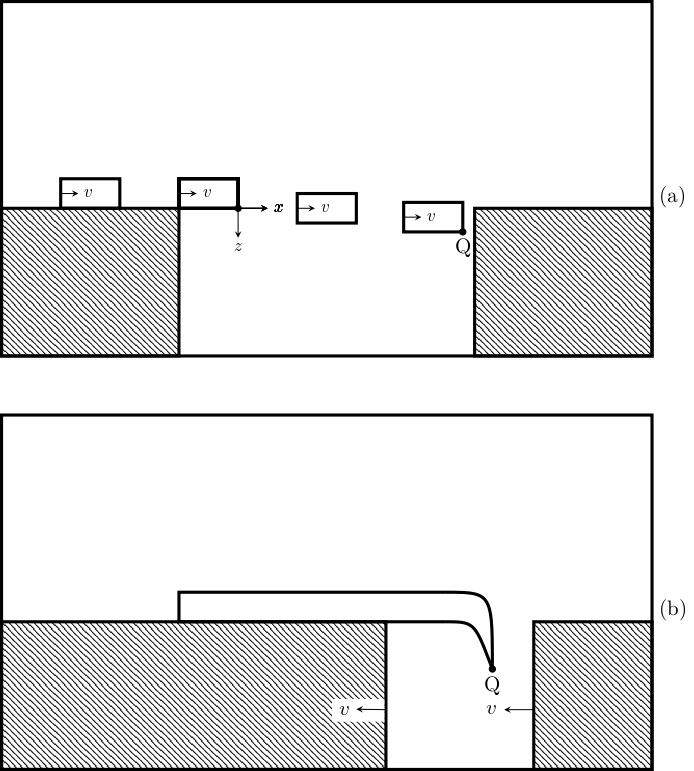}
    \caption{Queda da barra horizontal nos referenciais \textit{S} (a) e \textit{S'} (b).}
    \label{fig:barra}
\end{figure}

Seja P a extremidade traseira da barra e Q a extremidade dianteira (Fig. \ref{fig:barra}). Toma-se como origem comum $x = x'= 0$ a posição de Q no momento $t = t' = 0$ em que o alçapão se abre. Em \textit{S}, a barra estará contraída e inteiramente sobre o vão; em \textit{S'}, o vão estará contraído e apenas uma parte dela estará sobre ele, conforme a figura. Assim, aplicam-se as transformações:
\begin{equation}
    z = z', \hspace{0.5cm} t = \gamma\left(t' + \frac{vx'}{c^2}\right)
\end{equation}
No referencial \textit{S}:
\begin{equation}
    z = 0 \hspace{0.3cm} para \hspace{0.3cm} t < 0, \hspace{0.5cm} z = \frac{gt^2}{2} \hspace{0.3cm} para \hspace{0.3cm} t \geq 0
\end{equation}
Utilizando as transformações, tem-se as coordenadas em \textit{S'}:

\begin{gather}
 z' = 0 \quad para \quad t' < -\frac{c^2x'}{2}, \\
z' = \frac{g\gamma^2}{2}\left(t' + \frac{vx'}{2}\right)^2 \quad para \quad t' \geq-\frac{c^2x'}{v} 
\end{gather}
 
Pode-se interpretar a trajetória da barra em \textit{S'} como uma parábola cuja distância do foco até o vértice é $2c^4/g\gamma^2c^2$, com vértice inicialmente em Q. À medida que a extremidade esquerda do vão viaja para a esquerda, ela "puxa" o vértice da parábola e a barra se distorce. Em S', a barra deixa de ser rígida.
\\
\\
É fácil observar que, em \textit{S}, a extremidade esquerda do vão está inicialmente em $x_1=-L/\gamma$, e a extremidade direita, em $x_2=L(\gamma-1)/\gamma$. Assim, utilizando a contração de Lorentz, suas coordenadas iniciais em \textit{S'} são:
\begin{equation}
    x'_1=\gamma x_1=-L/\gamma^2, \hspace{0.5cm} x'_2=\gamma x_2 = L(\gamma - 1)/\gamma^2
\end{equation}
Seja $t_1$ o momento em que P atravessa a extremidade esquerda do vão e $t_2$ o momento em que Q toca a parede da direita. Em \textit{S}, é visível que $t_2 > t_1$; contudo, em \textit{S'}, a ordem dos eventos se inverte[10].
\begin{equation}
    0 = L(\gamma - 1)/\gamma^2 - vt'_2 \ \Rightarrow \ t'_2 = L(\gamma - 1)/v\gamma^2
\end{equation}
Em $t'_2$, a coordenada da borda esquerda será:
\begin{equation}
    x'_1 = -L/\gamma^2 - L(\gamma - 1)/\gamma^2 = -L/\gamma
\end{equation}
$x'_1 < -L$, significando que P ainda não está sobre o vão ($t'_2 < t'_1$). Se a distância entre as paredes do vão se mantém constante em \textit{S'}, a posição da parede da direita em $t'_1$ será $-L(\gamma-1)/\gamma < 0$, indicando que a barra sofrerá uma contração horizontal em \textit{S'} a partir da extremidade Q devido à colisão.

\subsubsection{Solução de Sartori}
Uma versão desse paradoxo é qualitativamente discutida em \textit{Understanding Relativity}, por Leo Sartori \cite{Sartori}. Nela, o alçapão em queda é substituído por cordas que suspendem a barra e são cortadas simultaneamente em \textit{S} quando a extremidade \textit{P} fica sobre o vão. O movimento será matematicamente idêntico à primeira versão do paradoxo, mas essa modificação torna bastante visível a razão da perda de rigidez da barra em \textit{S'}.
\\
\begin{figure}[h!]
    \centering
    \includegraphics[scale=0.92]{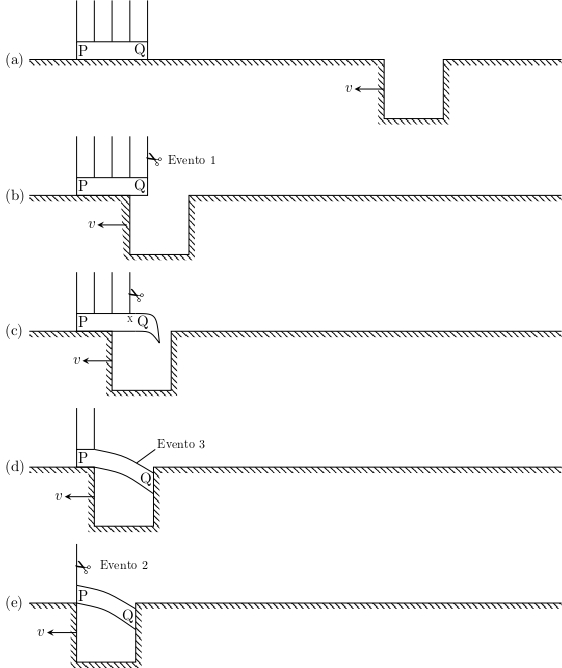}
    \caption{Corte das cordas visto do referencial \textit{S'}.}
    \label{fig:barra2}
\end{figure}
\\
O corte das cordas não é simultâneo em \textit{S'}, mas começa a acontecer na extremidade Q e "viaja" ao longo da barra até P. De forma semelhante ao que foi apontado na discussão sobre o paradoxo da guilhotina, pode-se pensar na barra como formada por vários pedaços verticais, cada um ligado a uma corda. No momento em que a corda em Q é cortada, o segmento sai do equilíbrio e começa a cair de forma independente dos outros; para que possa ser freado, é necessária a ação das forças de cisalhamento pelo restante da barra ainda em repouso.
\\

A informação de que Q está caindo, contudo, não atinge o segmento seguinte antes que as cordas dele sejam também cortadas, pois viaja abaixo da velocidade da luz (na prática, possui a velocidade do som na barra). O corte do próximo segmento sempre acontecerá antes porque todos os cortes são simultâneos em \textit{S}, indicando que o intervalo invariante entre cada um é de natureza \textit{espacial}:
\begin{equation}
    \Delta s^2 = (c\Delta t)^2 - (\Delta x)^2 < 0 
\end{equation}
Dessa forma, a separação entre segmentos adjacentes sempre será maior do que a luz poderia percorrer no intervalo correspondente; as forças de cisalhamento, portanto, nunca têm tempo de agir. Na prática, cada segmento entra em queda livre, formando a parábola anteriormente deduzida.
\\


\section{Naves espaciais de Bell}
\label{bell}
Este paradoxo foi introduzido por E. Dewan e M. Beran em 1959 \cite{Dewan.Beran}. Inicialmente tratado de forma qualitativa por Dewan \cite{Dewan} e Evett \cite{Evett}, o paradoxo das naves espaciais tornou-se mais conhecido após a discussão realizada por Bell \cite{Bell}, e desde então discussões a respeito de diversas propostas como as de Flores \cite{Flores}, Franklin \cite{Franklin}, Redžić \cite{Redžić} e Lewis \cite{Lewis.etal} vêm sendo publicadas até os dias atuais.

Suponha uma dupla de naves idênticas, onde estão Alice e Bob, respectivamente, inicialmente estacionárias entre si e em relação à uma terceira nave, onde está Charlie. As naves de Alice e Bob estão ligadas por uma longa corrente inextensível de comprimento próprio $L_{0}$. Seja $S$ o referencial de repouso de Charlie. No instante $t = 0$, conforme medido por Charlie, as naves de Alice e Bob ativam seus motores simultaneamente e começam a acelerar a uma taxa constante na mesma direção e sentido, que consideraremos a direção do eixo $x$, no sentido positivo. Charlie observa as naves de Alice e Bob acelerarem até atingir uma velocidade final $v$ em relação a ele, e então a aceleração de ambas termina simultaneamente. Como, para Charlie, as naves saem simultaneamente, a distância entre Alice e Bob ($x_{B}-x_{A}=L_{0}$) se mantém conforme ambas aceleram. Todavia, devido aos efeitos da contração de Lorentz, a corrente que conecta as naves de Alice e Bob se contrai. Assim, torna-se menor que $L_{0}$ e arrebenta. Seja agora $S'$ o referencial de Alice e Bob. Para eles,  a corrente esteve sempre em repouso, mantendo seu comprimento próprio $L_{0}$, e não deve arrebentar. Esse é o paradoxo das naves de Bell. A corrente arrebenta ou não?

A solução para este paradoxo está na simultaneidade. Portanto, assim como no caso da madeireira relativística \ref{madeireira}, a corrente arrebentar ou não dependerá de que eventos escolhemos serem simultâneos. Vejamos primeiramente o caso em que as acelerações comecem simultaneamente para Charlie. Como exposto acima, para ele a corda arrebenta. Vejamos o que ocorre para Alice e Bob. Se as naves saem simultaneamente para Charlie, temos $\Delta t=0$, e 
\begin{equation}
    \Delta t' = \gamma(v)\left [ \Delta t - \frac{v\Delta x}{c^2} \right]=-\frac{\gamma(v)L_{0}v}{c^2}
\end{equation}
 Em $S'$, os eventos A e B, o início da aceleração das naves de Alice e Bob \textbf{não} são simultâneos. A falta de simultaneidade para Alice e Bob se manifesta com a nave de Bob começando a acelerar enquanto a nave de Alice ainda está parada, resultando na ruptura da corrente que as conecta.
 
 Podemos agora pensar na situação em que a saída seja simultânea para Alice e Bob. Nesse caso, obviamente, a corrente nunca arrebenta para eles. Entretanto, a ausência de simultaneidade agora irá se manifestar para Charlie. Para ele teremos 
\begin{equation}
    \Delta t = \gamma(v)\left [ \Delta t' +\frac{v\Delta x'}{c^2} \right]=\frac{\gamma(v)L_{0}v}{c^2}.
\end{equation}
Ou seja, a nave de Alice começa a acelerar antes. Dessa forma, do ponto de vista de Charlie, a distância entre as naves deve se reduzir à medida que elas aceleram. Ora, isso é exatamente o esperado para que a corrente não arrebentem, pois teremos a contração de Lorentz da mesma. Isto é, a corrente se contrai, mas a distância entre as naves diminui na mesma proporção. 

Finalmente, o paradoxo de Bell merece algumas discussões adicionais. Considere o primeiro dos casos acima. Um questionamento comum é o seguinte: por que o comprimento da corrente sofre contração de Lorentz mas a distância entre as naves não, levando, portanto, à quebra da corrente? O ponto é que, em $S'$, o comprimento da corrente é \textbf{definido} simultaneamente. Ao definirmos que a naves partem simultaneamente para $S$, geramos o problema exposto acima: para $S'$, a nave dianteira parte antes. Isso levanta pontos bastante interessantes sobre aceleração de corpos rígidos. Imagine que, em vez de cordas e naves, consideremos uma única nave, com Alice na extremidade esquerda e Bob na direita. Ao acelerar, tensões podem romper um corpo rígido, no caso a nave? Isso é discutido em detalhes no livro do W. Rindler \cite{RindlerLivro}. 

Imagine que queremos acelerar um corpo rígido de tal forma que, para os sucessivos referenciais de repouso, ele mantenha seu comprimento próprio $L_0$. Ou seja, para Alice e Bob a nave sem tem seu comprimento próprio. A única forma de obter isso é que cada extremidade siga a trajetória \cite{RindlerLivro}
\begin{equation}\label{acelera}
    x^2-c^2t^2=x^2_0=c^4/\alpha^2.
\end{equation} 
Na equação acima, $x_0$ é a posição inicial e  $\alpha$ a aceleração própria. Como as posições iniciais de Bob e Alice são diferentes, a Eq.(\ref{acelera}) implica que as extremidades devem ter acelerações próprias diferentes. Portanto precisamos que, mesmo para Bob e Alice, além de partir simultaneamente, as acelerações sejam diferentes! Dessa forma, a nave segue sem tensões e sempre com o mesmo tamanho para Alice e Bob. Para Charlie, a corpo se contrairá, sucessivamente,  exatamente pelo fator de Lorentz \cite{RindlerLivro}. 

\section{Paradoxo das tesouras}
\label{tesouras}
Esse paradoxo, que aparece na literatura desde 1960, envolve a possibilidade aparente de velocidades superluminais \cite{Rothman}. Em anos seguintes, o aparente paradoxo foi abordado por S. Chase \cite{Chase} e, mais recentemente, vem sendo discutido por N. Kaushal, R.J. Nemiroff \cite{Kaushal.Nemiroff} e D. Faccio \cite{Faccio.etal} em uma série de artigos, cuja relevância se estende para além de problemas clássicos de da relatividade restrita, chegando à astronomia, onde ocorre a observação de uma classe de fenômenos de aparência superluminal \cite{Nemiroff}.

\begin{figure}[h]
    \centering
    \includegraphics{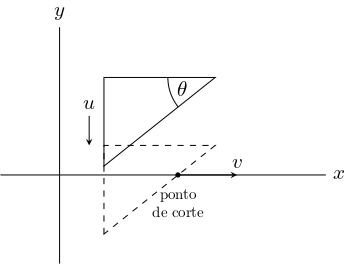}
    \caption{Guilhotina rígida.}
    \label{fig:lamina2}
\end{figure}

Imagine uma lâmina que faça um angulo $\theta$ com um papel, como na Fig. \ref{fig:lamina2}. À medida que desce, o corte pode atingir uma velocidade maior que a da luz. Para isso, basta imaginar uma situação em que a lâmina está na horizontal. Nesse caso, a velocidade se torna infinita, pois os pontos são cortados simultaneamente. 
Diferente dos casos anteriores, a solução para esse paradoxo não envolve as transformações de Lorentz. Basta notar que o ponto de corte não é um ponto material que se move propriamente pelo espaço. Cada ponto em que as lâminas se cruzam pode ser considerado um evento que ocorre localmente, sem interferir nos demais. Não havendo ligação causal entre eles, não há transporte de matéria nem energia acima da velocidade da luz. 

O paradoxo acima aparece em diversos livros de relatividade. Ele se torna interessante na medida em que leva a conclusões interessantes sobre os limites impostos ao movimento de objetos com  massa e informação através do espaço-tempo. Aqui daremos algumas versões dele, de acordo com o descrito por Kaushal e Nemiroff  \cite{Kaushal.Nemiroff}. 

Primeiramente vejamos o caso da Fig. \ref{fig:lamina2}. A lâmina da guilhotina cai verticalmente sobre a lâmina estática com velocidade constante $u$, e o ângulo entre ambas é $\theta$. O ponto de corte se move ao longo do eixo $x$ em velocidade constante $v$. Por geometria simples, obtemos que a relação entre essas grandezas é dada por:
\begin{equation}
    \label{eq:0201}
    v=u\, \cot \theta 
\end{equation}
Substituindo $v$ por $c$ na Eq.(\ref{eq:0201}), obtemos que a condição necessária para que a velocidade do ponto de corte atinja a velocidade da luz é dada por:
\begin{equation}
    \label{eq:0202}
    \theta _{c}=\textrm{arccot}\left(\frac{c}{u}\right)
\end{equation}

\begin{figure}[h]
    \centering
    \includegraphics{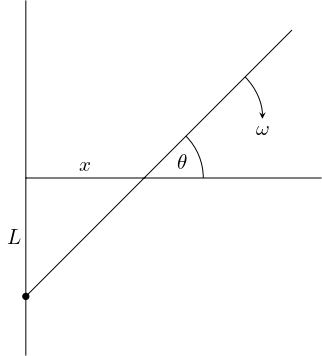}
    \caption{Lâmina rígida em rotação.}
    \label{fig:lamina1}
\end{figure}

A segundo caso é o de um par de lâminas infinitamente longas e finas como um par de retas em um plano cartesiano, como na figura \ref{fig:lamina1}. Uma delas é estática e permanece ao longo do eixo $x$ do plano, e a outra é móvel. Esta última cruza a lâmina estática com um certo ângulo $\theta$ e o eixo $y$ no ponto $(0,-L)$, sendo que ela gira em torno deste ponto, e a distância $L$ entre o ponto de rotação da lâmina e a origem definimos como "distância da dobradiça". O ponto em que as lâminas se cruzam definimos como o "ponto de corte", e a distância entre a origem do sistema e o ponto de corte em determinado instante é $d$.

Podemos imaginar a situação inicial, em $t = 0$ e $d = 0$, em que a lâmina móvel está paralela ao eixo $y$, portanto perpendicular à outra lâmina e começa a girar em sentido horário, fechando-se sobre ela. O ponto-chave deste caso é considerar as lâminas da tesoura perfeitamente rígidas, assim a lâmina móvel cruza a outra com uma velocidade angular constante $\omega$. Nesta situação, temos que a distância $d$ é:
\begin{equation}
    \label{eq:0101}
    d = L\, \cot \theta 
\end{equation}
Derivando a Eq.(\ref{eq:0101}) em relação ao tempo, verificamos que a velocidade do ponto do corte $v$ é dada por:
\begin{equation}
 \label{eq:0102}
    v = \omega L\, \csc ^{2}\theta 
\end{equation}

 Com $\omega$ e $L$ constantes, $v$ cresce muito rapidamente conforme o ângulo $\theta$ entre as lâminas da tesoura diminui, e eventualmente, ao se fecharem, as lâminas tornam-se paralelas e a distância $d$ infinita. Como isso ocorre em intervalo de tempo finito, a velocidade do ponto de corte cresce ilimitadamente, eventualmente ultrapassando a velocidade da luz, uma situação aparentemente paradoxal. Substituindo $v$ por $c$ na Eq.(\ref{eq:0102}), obtém-se que o ângulo, o instante e a distância horizontal em que a velocidade do ponto de corte se tornaria superluminal são dados respectivamente por:
 \\
 \begin{equation}
    \label{eq:0103}
    \theta _{c}=\arcsin \sqrt{\frac{\omega L}{c}};\,       t_{c}=\frac{(\frac{\pi }{2})-\theta _{c}}{\omega }; \,d_{c}=\sqrt{\frac{L}{\omega }(c-\omega L)}.
 \end{equation}

Podemos ressaltar que nesta situação, para quaisquer $\omega>0$ e $L>0$, o limite superluminal sempre será atingido.

\section{Conclusão}
Ao estudar a Relatividade Especial (RE),  deve-se abandonar a noção de espaço-tempo absoluto, descrito pelas transformações de Galileu. Todavia, mesmo após um primeiro contato com a RE,  os conceitos Galileanos ainda se fazem arraigados em muitos estudantes. Isso pode ser notado na dificuldade que os mesmo tem em resolver os paradoxos da relatividade. O fato é que, com esses paradoxos, os estudantes são confrontados com situações aparentemente contraditórias mas que, de fato, demonstram a  dificuldade em se adequar ao novos conceitos de espaço e tempo da RE. Com objetivo de preencher essa lacuna, esse artigo apresenta detalhadamente a solução de diversos desses paradoxos. 

Particularmente importante é a ausência de simultaneidade. Na seção \ref{gemeos}, estudamos o paradoxo dos gêmeos. Com a solução apresentada, podemos  afirmar que o aparente paradoxo se deve à dificuldade de abandonar o conceito Galileano de simultaneidade. O mesmo pode ser dito sobre os paradoxos da madeireira na seção \ref{madeireira}, das naves de Bell na seção \ref{bell} e do fenda e barra sob efeito da gravidade, seção \ref{barraefendagrav}. Esses três últimos podem ser inteiramente resolvidos ao se abandonar a simultaneidade absoluta. 

Abordamos ainda  paradoxos que envolvem o fato de $c$ ser uma  velocidade limite de propagação de qualquer sinal.  Isso coloca em xeque o conceito Galileano de corpo rígido. Abandonar isso foi a forma de resolver os paradoxos da guilhotina emperrada na seção \ref{emperrada} e da chave na seção \ref{chave}. Na seção \ref{tesouras}, abordamos o aparente paradoxo de propagações superluminais.

Finalmente, na seção \ref{barraefenda}, abordamos o paradoxo da fenda e barra sem efeito da gravidade.  Esse caso é particularmente interessante pois difere dos paradoxos unidimensionais: Além da simultaneidade, devemos abandonar a ideia  de que ângulos são iguais. No caso extremos vimos como, no limite ultrarelativístico, a fenda se inclina até um angulo reto.  


\section*{Agradecimentos}

We acknowledge the financial support provided by the Conselho Nacional de Desenvolvimento Científico
e Tecnológico (CNPq) and Fundação Cearense de Apoio ao Desenvolvimento Científico e
Tecnológico (FUNCAP) through PRONEM PNE0112- 00085.01.00/16.

\end{document}